\documentclass[article,twocolumn,showpacs,superscriptaddress,amsmath,amssymb]{revtex4}
\usepackage{cancel}
\usepackage{maybemath}
\usepackage{xcolor}
\usepackage{graphicx}

\begin{document}
Pub. in Lett. in High En. Phys., LHEP-199, 2021.\\http://journals.andromedapublisher.com/index.php/LHEP/article/view/199\\
\title{5 reasons to expect an 8 MeV line in the SN 1987A neutrino spectrum}
\author{Robert Ehrlich}
\affiliation{George Mason University, Fairfax, VA 22030}
\email{rehrlich@gmu.edu}
\date{\today}

\begin{abstract}
Evidence was previously reported for an 8 MeV neutrino line associated with SN 1987A based on an analysis of 997 events recorded in the Kamiokande-II detector on the day of the supernova.  That claimed line, however, occurred at the peak of the background spectrum, and both had a similar shape, making the claim tenuous at best.  Here the claim is buttressed by providing five reasons to expect such an 8 MeV neutrino line.  A final section of the paper concerns the ongoing KATRIN experiment to find the neutrino mass, which might provide additional support for the line, should it validate a controversial $3+3$ model of the neutrino masses, including a tachyonic ($m^2<0$) mass.\\

ArXiv/2101.08128\\

keywords: neutrino line, monochromatic neutrinos, SN 1987A, supernova, dark matter, galactic center, KATRIN experiment
\end{abstract}
\maketitle

\section{Introduction}

Elsewhere evidence was presented for an 8 MeV neutrino line in the SN 1987A spectrum -- see appendix in ref.~\citep{Eh2018}.  The line had the right shape and width, and its background was independently derived.  One potentially fatal problem, however, was that the line occurred at the peak of the background and had a similar shape, making its existence uncertain.  This paper buttresses the claim by providing five reasons to expect such an 8 MeV line from SN 1987A.  In addition, it presents other new evidence that makes the case for the line stronger, and discusses how potential contradictions can be satisfactorily addressed.  Finally, two specific tests are proposed for such a neutrino line.  One test involves searches for diffuse supernovae using a novel approach with existing data, and the second involves the ongoing KATRIN experiment to measure the electron neutrino mass.  

The data in support of an 8 MeV neutrino line are the 997 events recorded on the date of SN 1987A by the Kamiokande-II detector~\citep{Hi1988}.  Before reviewing that data, let us first explain three of the five reasons to expect such a supernova neutrino line.  They all involve the possible existence of cold dark matter particles of mass $m_X\sim 10 MeV.$

\section{1st Reason: Galactic center $\gamma$-rays}
The galactic center (GC) has been considered as a possible place for dark matter (DM) annihilations to occur, in which case the $\gamma-$rays from the GC could be the result of $XX\rightarrow e^+e^-$ followed by $e^+e^-\rightarrow 2\gamma.$  We can, therefore, learn about the possible presence of DM near the GC by examining the spectrum of $\gamma-$rays from that source.  There is also a direct connection between $XX$ annihilations and neutrino lines.  Thus, if there were evidence for cold dark matter particles of mass $m_X,$ their annihilation $XX\rightarrow \nu  \bar{\nu},$ would give rise to  be nearly monochromatic $\nu, \bar{\nu}$ having $E=m_X$.  

Fig. 1 shows the predicted enhancement above background for the GC $\gamma-$ ray spectrum due to dark matter annihilation.  The four curves correspond to different $m_X$ values.  These curves are found by assuming that the $e^+$ are created in $XX\rightarrow e^+e^-$ with an initial energy $E_0=m_X.$  Of those $e^+,$ we assume $97\%$ will annihilate at rest yielding the 511 keV line, while the remaining $3\%$ propagate in a neutral medium before annihilating in flight~\citep{Je2006}.  Also shown in the figure is the GC $\gamma-$ ray flux data from four instruments.  Note that most of the data and three of the four enhancement curves previously appeared in refs.~\citep{Si2006,Pr2011}, but the author has added the 8.3 MeV enhancement curve and the 7 OSSE points from Ref.~\citep{Ki2001}.  

The data in Fig. 1 can be seen to be consistent with $m_X=10$ MeV (black curve), with $\chi^2=7.3,$ $p=89\%, dof=13.$   In contrast, the fit to the null hypothesis, i.e., the dashed line power law, is completely unacceptable: $\chi^2 = 960.$  Acceptable fits to the data in Fig. 1 can only be found for the range: $m_X=10^{+5}_{-1.7}$ MeV.  Thus, the null hypothesis is excluded by $N=10/1.7\sim 6$ standard deviations.  This rejection of the null hypothesis is in marked contrast to the conclusion in refs.~\citep{Si2006,Pr2011}, which failed to include the OSSE data.  The key role of the OSSE data here arises from their very small error bars, which is discussed in section 3.1.3 of ref.~\citep{Eh2018}.

We now consider other data that strengthens the case for $m_X\sim 10 MeV,$ and hence a neutrino line with this energy.  Incidentally, an $m_X\sim 10 MeV$ for a DM particle is just within BBN and CMB cosmological constraints, which excludes a thermal dark matter
particle with a mass $m_X< 7–10 MeV.$~\citep{De2019}. 

\begin{figure}
\centerline{\includegraphics[angle=0,width=1.0\linewidth]{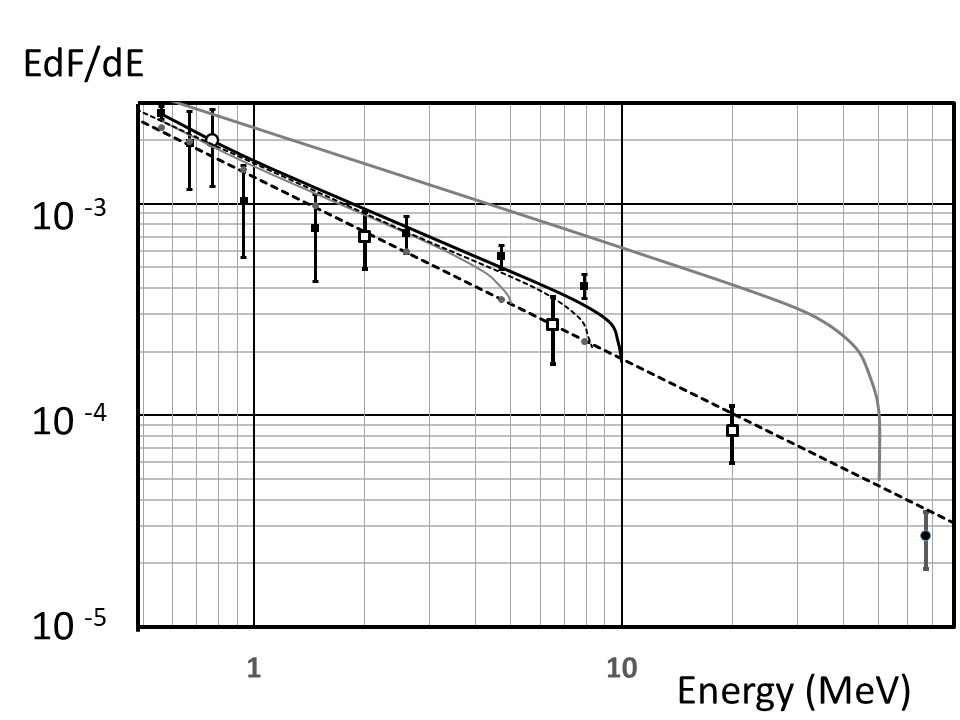}}
\caption{Flux, i.e., $E\times\frac{dF}{dE} (cm^{-2}s^{-1})$ versus energy for $\gamma-$rays from the inner galaxy, as measured by: SPI(open circle), COMPTEL (open squares), EGRET (filled circles), and OSSE (filled triangles).  All but the OSSE data (from Ref.\citep{Ki2001}) are from ref.~\citep{Si2006}.  The computed enhancements above the straight line are for positrons injected into a neutral medium at initial energies $E_0=m_X=5, 8.3, 10, 50$ MeV displayed respectively as: lower grey, dotted, black, and upper grey.  The sloped line  is a power law (index 1.55) fit to data at high and low energies.\label{fig1}}
\end{figure}

\section{2nd reason: the $Z'$ Boson}
If cold dark matter $X$ particles having $m_X\sim 10 MeV$ exist and their annihilation yields monochromatic $e^+e^-$ pairs as suggested in the previous section, it is reasonable to suppose that the reaction proceeds via some mediator particle Z' as in $XX\rightarrow Z'\rightarrow e^+e^-,$ whose mass is $m_{Z'}= 2 m_X,$ by energy conservation. The natural place to look for such a Z' would be in a nuclear physics experiment, where the decay of some nuclear excited state $N^*$ produced $e^+e^-$ pairs via $N^*\rightarrow N +Z'\rightarrow N+e^+e^-.$  Of course, most of the time when $e^+e^-$ pairs are observed it would be when the mediator particle is a photon, so the existence of such a Z' would be revealed by an enhancement to that reaction, i.e, an excess of $e^+e^-$ pairs having a specific opening angle, corresponding to $m_{Z'}.$  In 2016 exactly such an enhancement was reported by Krasznahorkay et al. (the Atomki group) for $e^+e^-$ emissions in the reaction $^7Li(p,\gamma)Be^8.$~\citep{Kr2016}  Their result implied an intermediate short-lived Z' particle (sometimes called X17) with mass $m=16.7\pm 0.6$ MeV appearing in the two step decay process of the excited $^8Be,$ i.e.: $^8Be^*\rightarrow ^8Be Z',$ followed by $Z'\rightarrow e^+e^-.$  

In 2020 the Atomki group has reported the same anomaly in the decay of excited helium atoms in the reaction $^3H(p,\gamma)^4He$ as they earlier observed in $^8Be$.~\citep{Kr2020}.  
A particle having a 16.7 MeV mass would also be expected to be found in some accelerator experiments.  However, the NA64 experiment (and others) at the CERN SPS have not observed it.~\citep{De2020}.  On the other hand, these negative results do not contradict those in refs.~\citep{Kr2016,Kr2020} because they were not sensitive to a small range of particle lifetimes consistent with that reported in ref.~\citep{Kr2016} -- see Fig. 1 in ref.~\citep{De2020}.  

Various discrepancies from standard model predictions also support the Z' interpretation of the Atomki anomaly.  Thus, ref.~\citep{Ki2020} explains how the $(g - 2)_\mu$ anomaly $(3.7\sigma)$can be explained based on an extension of the Standard Model, including a light Z' boson as observed by the Atomki group.
In addition, the neutron lifetime puzzle can be explained by assuming a virtual $Z'$ exchange into a neutrino and its Kaluza-Klein sibling.\citep{Du2020} Finally, such a particle could account for the $\sim 2-3\sigma$ deviations from standard model predictions seen in the leptonic decays of the $\pi$~\citep{Al2020} and $B$~\citep{Gr2019} mesons.  

However, the $Z'$ boson interpretation of the Atomki anomaly has also been challenged.  For example, Aleksejevs etal.~\citep{Al2021} and Koch~\citep{Ko2020}, claim that the observations can be explained within the standard model by (1) adding the full set of second-order corrections and the interference terms to the Born-level decay amplitudes, and (2) accounting for detector and analysis bias.  Similarly, Zhang and Miller also provide an alternate standard model explanation of the Atomki anomaly.~\citep{Zh2021}.  However, refs.~\citep{Al2021,Ko2020,Zh2021} only apply to the $^8B$ data and they did not consider the similar anomaly seen in the helium case.  Furthermore, the analysis in ref.~\citep{Zh2021} only rules out a new $Z'$ {\emph{vector}} boson as the explanation of the Atomki anomaly, not a scalar.

If the anomalies observed in refs.~\citep{Kr2016, Kr2020} really are due to a new $Z'$ boson, this particle would be the mediator of a fifth force.~\citep{Fe2016}  Moreover, as already noted, for cold DM $X$ particles, the end product of $XX\rightarrow Z'\rightarrow \nu\bar{\nu}$ (which is the only other Z' decay mode according to ref.~\citep{Ch2016}) would be nearly monochromatic $\nu$ and $\bar{\nu}$ pairs having $E_\nu=8.4\pm 0.3$ MeV.  

\section{3rd reason: DM in supernovae}
\label{supernova}

Many researchers have suggested that dark matter might collect in the core of some stars.~\citep{Ba2008}  DM annihilation triggering a supernova is plausible because without such an ``extra" energy, shock wave stalling has been a difficulty with most supernova models, the best of which have elements that can still only be understood in qualitative terms.~\citep{Ja2017b}  
It may be true that as of 2020 a self-consistent 3D simulation with detailed neutrino transport has finally achieved a neutrino-driven explosion with properties similar to SN 1987A without dark matter.~\citep{Bo2020}  However, even if DM may not be required to trigger a neutrino-driven explosion, the presence of large amounts of DM in the stellar core could still play a role in the explosion and be the source of significant long-lasting monochromatic neutrino emissions.  In fact, Fayet et al.~\citep{Fa2006} have shown that $m_X\sim1-30 MeV$ dark matter particles can play a significant role in core-collapse supernovae. They also note that if the DM particles have relatively large annihilation and scattering cross sections, and have $m_X<10MeV,$ the DM would cool on a time scale perhaps $>100$ times that in the standard scenario,~\citep{Fa2006} as would be implied by an analysis of the Kamiokande data now described. 

\section{Analysis of Kamiokande data}
The largest of the four detectors operating at the time of the SN 1987A observation was Kamiokande-II.~\citep{Hi1988}  This detector recorded neutrino arrival times and their energies, which could be deduced from the ``visible"  energies, $E_{vis}$ based on $E_\nu=E_{vis}+1.3$ MeV, assuming the dominant reaction to be $\bar{\nu}_e+p\rightarrow n+e^+.$  In addition to observing the main 12-event burst, Kamiokande-II also recorded 997 events occurring during eight 17-min long intervals during several hours before and after the burst. Figs. 4 (a)-(h) of Ref.~\citep{Hi1988} show scatter plots  for each event displaying the number of ``hits," $N_{hit},$ (PMT's activated) versus the event occurrence time, $t,$ during that $\Delta t = 8\times 17\rm{min}=0.094$ day time interval.  

Those eight plots of $N_{hit}$ vs $t$ were digitized by the author who then counted the number of times various $N_{hit}$ values occurred.  The $N_{hit}$ frequency distribution is shown in Fig. 2.  Note that the $N_{hit}$ values are found to be proportional to the visible energy, i.e., $E_{vis}=cN_{hit} MeV$ with $c=0.363\pm 5\%,$ as shown in Fig. 4(a) of ref.~\citep{Eh2018}.  Thus, Fig. 2 is actually a spectrum for the events observed over several hours, with the peak at 17 hits corresponding to $E_\nu=7.5\pm 0.5 MeV$.

\begin{figure}[t!]
\centerline{\includegraphics[angle=0,width=1.1\linewidth]{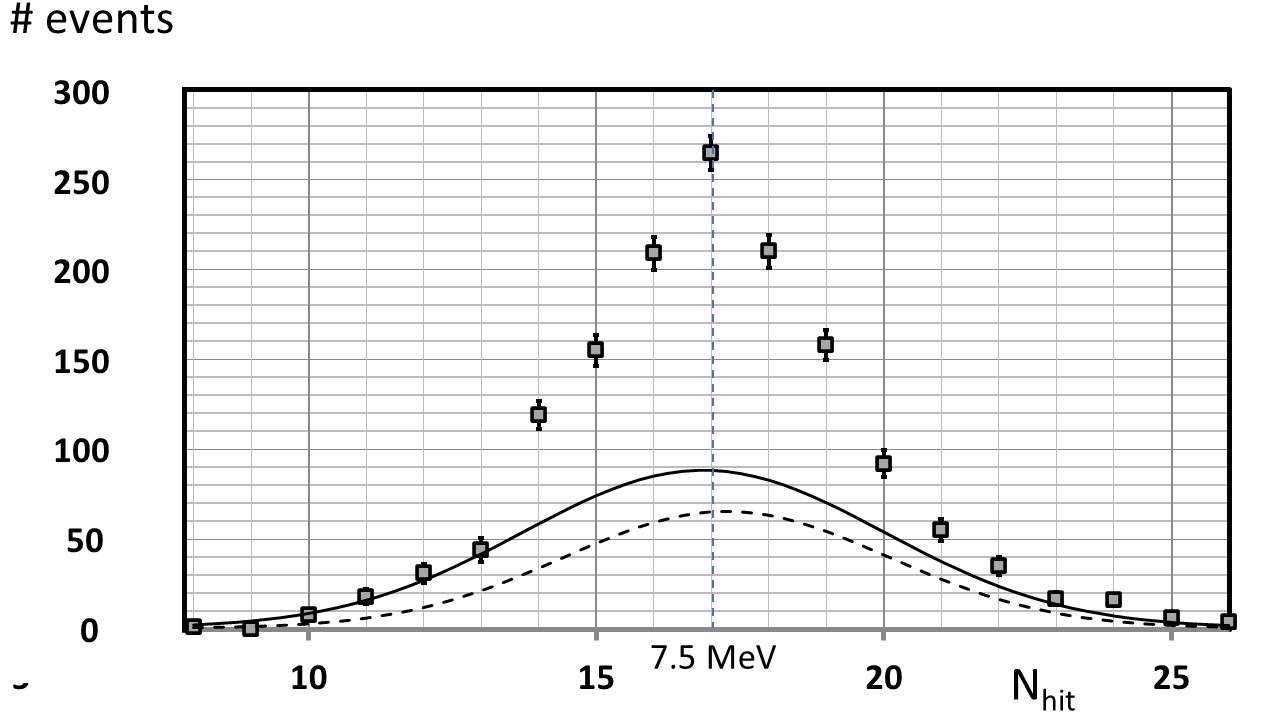}}
\caption{Kamiokande II neutrino data on Feb. 24, 1987.  Histogram of $N_{hit}$ values for 997 events in Fig. 4 in ref.~\citep{Hi1988} The solid and dashed curves are two versions of the background for the detector.  The dashed one was extracted from published data on a search for $^8B$ solar neutrinos -- see Appendix A in ref.~\citep{Eh2018} for details.  The solid background curve is based on a fit to the data using a Gaussian for the ten data points with $N_{hit}< 14$ and $N_{hit}> 22.$  The value $N_{hit}=17$ corresponds to $7.5\pm0.5$ MeV.\label{fig2}}
\end{figure}

\subsection{Finding the background spectrum}
Clearly, any claim of a peak above background in Fig. 2 depends critically on how the background is determined.  If the background is correctly represented here the peak would have very high statistical significance $(\sim 30\sigma),$ but no such claim can be made here given the strong similarity between the shape of the signal and background.  Two versions of the background are depicted in Fig. 2 -- one dashed and on solid.  The dashed background has been found based on a 1989 publication by the Kamiokande-II Collaboration (K-II) on a search for solar neutrinos from the reaction ${^8}B\rightarrow {^8Be^*}+e^++\nu_e.$~\citep{Hi1989}, as explained in Appendix A of ref.~\citep{Eh2018}.  The beginning of the 450 day data-taking period preceded the date of SN 1987A, but most of it was many months afterwards.  The excess counts above the dashed background in Fig. 2 can be well-fit by a Gaussian curve centered on $E_{max}=7.5\pm 0.4$ MeV, with a width consistent with the expected $25\%$ energy resolution based on $\Delta E/E=22\%/\sqrt{E/10},$~\citep{Hi1988} consistent with what would be expected for an 8 MeV neutrino line.

The most obvious flaw with using the data from the $^8B$ neutrino search for the background on the day of SN 1987A  would be if the background count rate in the detector were significantly time dependent.  It could be risky to assume a constant background over time given various improvements made to the detector.~\citep{Hi1991}  Therefore, the solid background curve in Fig. 2 has been found using the data themselves by the procedure outlined in the caption.  Of course, normally, when this technique is used to find evidence for a signal above an unknown background, the signal and background curves have very different shapes, and the evidence for the spectral line's existence would be much more certain than in the present case.

\section{Possible contradictions}
Here we discuss three possible contradictions to there being an 8 MeV neutrino line from SN 1987A, and show how they can be addressed.  

\subsection{Lack of time variation in line prominence}
If the 8 MeV line is real, one might have expected that of the eight 17-min time intervals for which ref.~\citep{Hi1988} provided data, those time intervals closest to the 12-event burst would show a greater excess above background, whereas no such variation is found in the data.  Unfortunately, there is no surviving record of the background on the days preceding and following the supernova.  The lack of any such time variation in the strength of the line is worrisome, but it would be consistent with the long-lasting luminosity expected for neutrinos emitted via light dark matter annihilation.~\citep{Fa2006}  If the DM annihilation in supernovae were strongly coupled to neutrinos, with neutrinos and light DM particles decoupling at $<\sim 3.3 MeV$ rather than the usual 8 MeV for neutrinos, the supernova cooling time scale would be larger by a factor of $\sim 20$ than in the standard scenario.~\citep{Fa2006}  Furthermore, in the interior of the interior of the proto-neutron star, the cooling time scale would be a factor $\ge 100$ larger~\citep{Fa2006}.  Such long cooling times would be expected to be matched by long heating times for the dark matter, so it would not be surprising to find that the emissions from DM annihilation are not simply confined to times after the main 12 event burst.  The neutrino emission of massive stars before a supernova has been studied by many authors and they all agree on significant, potentially detectable neutrino luminosities, see the review in Kato et al.~\citep{Ka2017}

\subsection{Non-observation in diffuse supernova searches}
The diffuse supernova neutrino background is a theoretical population of neutrinos cumulatively originating from all of the supernovae events which have occurred throughout the Universe for which only upper limits currently exist.   Given the large number of excess events above background comprising the 8 MeV neutrino line, one might expect it would have been detected in previous searches for diffuse supernovae.  However, those searches either focused on neutrinos having energies well above 8 MeV~\citep{SK2012,Ah2020} or alternatively neutrinos emitted in seconds-long bursts.~\citep{No2017}  Thus, despite those negative results, it might be possible to see such emissions even with today's neutrino detectors, especially those which have operated over many years, and have a low energy threshold.  In doing such a search one might look for a significant excess of counts in any day-long time interval for events in an energy band centered on $E_\nu=8 MeV.$  One day would be about the optimum size search window to use, because if the duration of the monochromatic emissions was much shorter than that, a time variation in the strength of the 8 MeV neutrino line would have been seen in the Kamiokande-II data, and if the monochromatic emissions lasted much longer than a day, the precursor star to SN 1987A would have to contain an impossible amount of dark matter -- see next section.

\subsection{Impossible number of 8 MeV neutrinos}
A third problematic aspect of the claimed 8 MeV neutrino line from SN 1987A is the sheer magnitude of the excess events in Fig. 2 being $\sim 700$ above background.  Recall that these data were from eight 17 minute long time intervals, i.e., roughly two hours duration chosen at random from a total time of ten hours.  Therefore, it is reasonable that given the lack of time variation seen over those ten hours, the total time during which monochromatic neutrinos were emitted by SN 1987A would need to be at least 12 hours duration.   In that case, one would estimate the number of monochromatic neutrinos in those 12 hours to be $700\times 12/2\sim 4,000,$ as compared to 12 events seen in the main burst.  In other words if we take the usual estimate of $10^{58}$ neutrinos emitted from SN 1987A, the number associated with DM annihilation would be $\sim 400$ times more or $4\times 10^{60}$.  Let us now see if such a value is even remotely possible.  

The star giving rise to SN 1987A was estimated to have 18 solar masses.  Let us assume it consisted of half DM, and that, as was assumed earlier, the DM X particles, had a mass of 8 MeV, yielding in total $\sim 10^{61}$ $X$ particles in the star.  We, therefore, find that at least a half the DM in the star would need to annihilate producing 8 MeV neutrinos to match the number $\sim 4\times 10^{60},$  of emitted monochromatic in an emission duration of half a day.  Thus, under our assumptions, finding as many as $\sim 700$ neutrinos constituting an observed 8 MeV spectral line in the Kamiokande II data is not impossible.  Lending further credence to the notion of such a large monochromatic neutrino emission, Brito et al.~\citep{Br2015} have shown that stellar cores  do not necessarily collapse when they grow beyond their Chandrasekhar limit, and that large fractions of DM in stellar cores are not ruled out.~\citep{Br2016}  
\\
\section{4th Reason: Mont Blanc Burst}
There are two more pieces of evidence in support of an 8 MeV neutrino line from SN 1987A in addition to the three previously noted.  First, the controversial Mont Blanc neutrino burst on the day of SN 1987A has been disregarded by most physicists with some exceptions~\citep{Gi1999,St1987,Vo1987}, owing to its 5 hour early arrival and its absence in the Kamiokande II detector.  Unlike the bursts seen in the other three detectors then operating which all had much higher energies, the five Mont Blanc neutrinos all have energies strangely consistent with the value 8 MeV within uncertainties,~\citep{Aga1988,Agb1987} in which case they could be the result of the 8 MeV neutrino line.   The absence 
of such a 5 hour early burst in the Kamiokande II data can be explained by the higher energy threshold of that detector ($20\%$ efficient at $E_\nu=8 MeV$),~\citep{Hi1988} and the synchronization of the two detectors being no better than $\sim \pm 1$ minute.  Aglietta et al. gives further reasons why the Kamiokande II detector would have probably seen only 1.5 neutrino events at the time of the Mont Blanc early burst.~\citep{Agc1987}.
\begin{figure}
\centerline{\includegraphics[angle=0,width=1.0\linewidth]{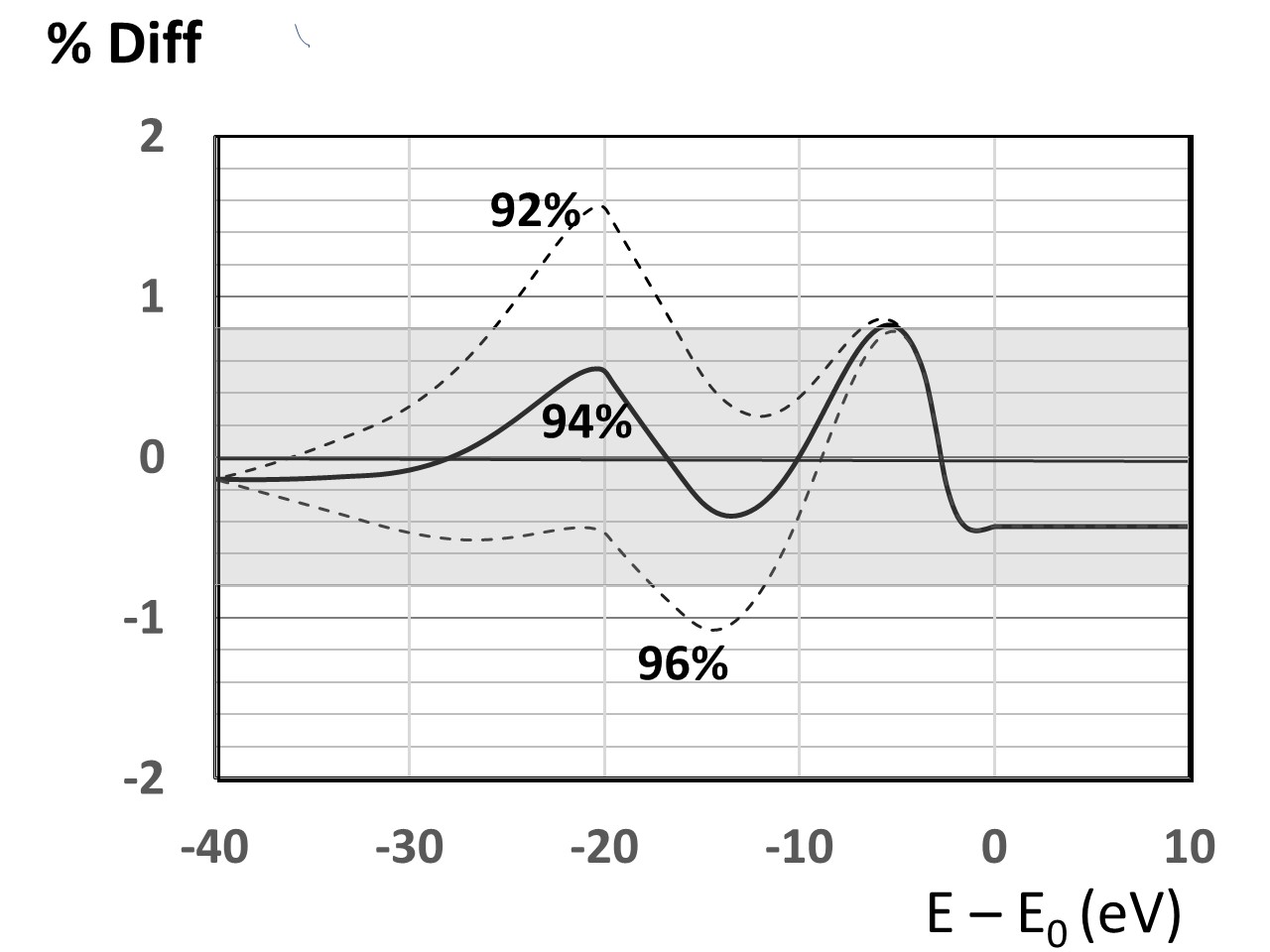}}
\caption{\label{fig1}Computed difference between the integral beta spectra for the $3+3$ model of the neutrino masses and the standard model of a small single effective mass.  For the $3+3$ model three values of the contribution for the 4.0 eV mass are shown: $92\%, 94\%$ and $96\%.$  Only for the $94\%$ case are the two integral spectra everywhere within $\pm 0.8\%$ of each other.  The graph shows the difference in the two integral spectra, because the integral spectrum is what KATRIN measures.}
\end{figure}

\section{5th Reason: $3+3$ model and KATRIN}
A final reason to expect there to be an 8 MeV neutrino line from SN 1987A is provided by the author's exotic $3+3$ neutrino model.  The model postulates that the neutrino flavor states are each comprised of three active-sterile doublets having these masses: $m_1=4.0\pm 0.5 eV,$ $m_2=21.4 \pm 1.2 eV,$ and $m_3^2\sim -0.2 keV^2.$~\citep{Eh2013}  Although the model is highly exotic, and involves a ``tachyonic" mass state, it has satisfied a number of empirical tests.~\citep{Eh2019a}  This controversial model requires any 5-hour early neutrino burst like that observed in the Mont Blanc detector to be monochromatic, with an energy roughly 8 MeV, something the author initially believed to be ``inconceivable."~\citep{Eh2013}   

The $3+3$ model can be tested in the KATRIN experiment, which is measuring the electron neutrino effective mass based on fitting the shape of the beta decay $\emph {integral}$ spectrum of tritium near its endpoint, $E_0.$  KATRIN's first results published in 2019~\citep{Ak2019} have been shown to yield a slightly better fit to the model than to a single small effective mass with $m_\nu<1 eV.$~\citep{Eh2019} However, a good fit to the model only is found for a narrow range of contributions from the 4.0 eV mass, specifically for $\alpha=|U_{i,1}|^2=0.94\pm 0.02.$\citep{Eh2019}  The need for a narrow range in $\alpha$ is illustrated in Fig. 3, which shows the computed difference in the $\emph {integral}$ spectra for the two models.  To find those curves, the differential spectrum for each mass $m_i$ is first found from the square of the Kurie function:
\begin{equation}
K_i^2(E)=(E_0-E)\sqrt{R[(E_0-E)^2-m_i^2]}
\end{equation}
where $R(x)$ is the ramp function ($R(x)=x$ for $x>0$ and $R(x)=0$ otherwise).  The respective integral spectra for a given mass is then found by a numerical integration of Eq. 1 from $E$ to $E_0,$ and the spectra for both signs of $m^2$ are assumed to vanish for $E-E_0>0.$ 

A consequence of Fig. 3 is that if the KATRIN data are well described by the $3+3$ model, but they are fit to a single small effective mass spectrum, then one would expect to see two ``bumps," that is, excess 
numbers of events ($O(0.5\%)$) at a distance from $E_0$ close to the two $m^2>0$ masses in the model.  The initial KATRIN results from 2019 were such that the statistical error bars were not small enough to clearly show whether these excesses were present, although the residuals for their best fit did show an indication they were.  In fact, the ninth residual in their fit located at $E-E_0 = -23 eV$, which was consistent with the position of the left peak in Fig. 3, fell $+2.5\sigma$ above their best four free parameter fit to a single effective mass.\citep{Eh2019}.  However, the most recent KATRIN experiment fit to the spectrum released in May 2021 has used not four free parameters to fit the data, but rather 37.~\citep{Ak2021}.  One of those parameters is the value of $m^2$ and the other 36 are the signal amplitude, $S_i$, background amplitude, $B_i,$ and endpoint energy, $E_{0,i},$ for the data recorded in each of 12 concentric rings on the detector.  While the assumption of a radial dependence of signal and background is reasonable in one sense, it was not done previously.  Moreover, the use of so many free parameters can mask the $O(0.5\%)$ departures from the single mass spectrum that would occur if the $3+3$ model were a valid description of the data -- see Fig. 3.  In fact, the number of adjustable parameters KATRIN uses in its fits is even greater than 37 if one counts some unspecified number of ``pull terms" corresponding to constrained parameters that can vary within limits.

The mathematician John Von Neumann famously once said: ``With four parameters I can fit an elephant, and with five I can make him wiggle his trunk."~\citep{Dy2004}  This quote was his humorous way of telling us to be suspicious of using too many free parameters in doing a fit.  Despite Von Neumann's caution, the use of four parameters by KATRIN to fit its initial data was exactly appropriate, but 37-plus parameters it now uses would seem to be excessive.   Whether the departures from the single effective mass spectrum indicated in Fig. 3 are in fact present in the KATRIN data remains uncertain at present, and will probably remain so, unless the number of free and partly free parameters used to fit the data is significantly reduced.  If four parameters are enough to fit an elephant, 37-plus are enough to fit (or here hide) a whole herd of elephants, all with their trunks wiggling.

\section*{Conflicts of interest}
I have no competing conflicts of interest.


\begin{thebibliography}{10}

\bibitem{Eh2018} R. Ehrlich, Astropart. Phys., {\bf99}, 21-29 (2018); arxiv.org/1701.00488.
\bibitem{Hi1988} K. S. Hirata, et al. (Kamiokande collaboration), Phys. Rev. D {\bf 38}, 448-458 (1988).
\bibitem{Je2006} P. J. Jean, et al., Astron. and Astrophys., {\bf 445}, 579 (2006).
\bibitem{Pr2011} N. Prantzos, et al. Rev. Mod. Phys., {\bf 83}, 1001 (2011).
\bibitem{Si2006} P. Sizun, M. Cassé, and S. Schanne, Phys. Rev. D {\bf74}, 063514 
\bibitem{Ki2001} R. L. Kinzer et al., ApJ, {\bf559}, 282, (2001).
\bibitem{De2019} P. F. Depta et al., JCAP 04, 029 (2019).
\bibitem{Kr2016} A. J. Krasznahorkay et al., Phys. Rev. Lett. {\bf 116}, 042501 (2016).
\bibitem{Kr2020} D. S. Firak et al., EPJ Web of Conferences {\bf232}, 04005 (2020).
\bibitem{De2020} E. Depero et al., Eur. Phys. J. C {\bf80} 1159 (2020). 
\bibitem{Ki2020} D. V. Kirpichnikov, V. E. Lyubovitskij and A. S. Zhevlakov, Phys. Rev. D {\bf102}, 095024 (2020).
\bibitem{Du2020} P. T. Du et al., J. Phys., Conf. Ser. 1506, 012004 (2020).
\bibitem{Al2020} D. S. M. Alves, Phys. Rev D, accepted, (2021); arxiv.org/2009.05578
\bibitem{Gr2019} E. Graverini, for the ATLAS, CMS and LHCb Collaborations, J. Phys. Conf. Ser. 1137, 012025 (2019).
\bibitem{Al2021} A. Aleksejevs, et al.; arxiv.org/2102.01127 (2021).
\bibitem{Ko2020} B. Koch, Nucl. Phys. A 1008, 122143 (2021).
\bibitem{Zh2021} X. Zhang and G. A. Miller, Phys. Lett. B {\bf 813}, 136061 (2021).
\bibitem{Fe2016} J. Feng et al., Phys. Rev. Lett. {\bf 117}, 071803 (2016).
\bibitem{Ch2016} C-S. Chen et al., Int. J. Mod. Phys. A {\bf32}, 1750178 (2017).
\bibitem{Ba2008} V. D. Barger, W. Y. Keung and G. Shaughnessy, Phys. Lett. B, {\bf 664}, 190 (2008).
\bibitem{Ja2017b} B. Muller, et al., MNRAS {\bf472} (1) 491–513 (2017).
\bibitem{Fa2006} P. Fayet, D. Hooper, and G. Sigl, PRL {\bf96}, 211302 (2006).
\bibitem{Bo2020} R. Bollig et al., submitted to ApJ, arXiv:2010.10506. 
\bibitem{Hi1989} K. S. Hirata, et al. (Kamiokande collaboration), Phys. Rev. Lett., {\bf 63}, 1 (1989).
\bibitem{Hi1991} K. S. Hirata, et al. (Kamiokande collaboration), Phys. Rev. {\bf44}, 8 (1991).
\bibitem{Ka2017} C. Kato et al., ApJ, {\b 848}, 48 (2017).
\bibitem{SK2012} The Super-Kamiokande Collaboration, K. Bays et al., Phys. Rev. D, {\bf85} (2012) 052007 [arXiv:1111.5031].
\bibitem{Ah2020} B. Aharmim et al. (SNO Collaboration), Phys. Rev. D, {\bf102}, 062006 (2020).
\bibitem{No2017} Yu. F. Novoseltsev, et al., J. of Exp. and Theor. Phys., {\bf125} (1) 73–79 (2017).
\bibitem{Br2015} R. Brito, V. Cardoso, and H. Okawa, Phys. Rev. Lett., {\bf 115}, 111301 (2015).
\bibitem{Br2016} R. Brito et al., Phys. Rev. D {\bf93}, 044045 (2016).
\bibitem{Aga1988} M. Aglietta et. al., Nucl. Phys. B (Proc Suppl.) {\bf 3} 453-462 (1988).
\bibitem{Agb1987} M. Aglietta et. al., Europhys. Lett., {\bf 3} (12) 1315 (1987).
\bibitem{Agc1987} M. Aglietta et al., Europhys. Lett., {\bf 3} (12), 1321-1324 (1987). 
\bibitem{Eh2013} R. Ehrlich, Astropart. Phys., {\bf41} (2013) 1–6; arxiv.org/1204.0484.
\bibitem{Eh2019a} R. Ehrlich, Adv. in Astron., 2820492 (2019); arXiv:1711.09897
\bibitem{Gi1999} S. Giani, STAIF-99 Proceedings, American Institute of Physics (1999).
\bibitem{St1987} L. Stella, and A. Treves, Astron. and Astrophys., {\bf185}, L5-L6 (1987)
\bibitem{Vo1987} D. N. Voskresensky, Astrophys. and Space Sci., {\bf138}, 421–424 (1987)
\bibitem{Ak2019} M. Aker et al., Phys. Rev. Lett. 123, 221802 (2019).
\bibitem{Eh2019} R. Ehrlich, Lett. in High En. Phys., {\bf2} (4), (2019); arxiv.org/1910.06158
\bibitem{Ak2021} M. Aker et al. (KATRIN Collaboration);\\arXiv:2105.08533.
\bibitem{Dy2004} F. Dyson, Nature, 427, 6972, 297 (2004).
\end{thebibliography}
\end{document}